\documentclass[twocolumn]{svjour3}
\smartqed
\usepackage{graphicx}
\usepackage{multirow}
\journalname{Granular Matter}
\begin{document}

\title{The Influence of the Degree of Heterogeneity on the Elastic Properties of Random Sphere Packings}

\author{Iwan Schenker \and Frank T. Filser \and  Markus H\"{u}tter \and Ludwig J. Gauckler}

\institute{I. Schenker (corresponding author) \and F. T. Filser
\and L. J. Gauckler \at
              Nonmetallic Materials, Department of Materials, ETH Zurich, 8093 Zurich, Switzerland\\
              \email{iwan.schenker@alumni.ethz.ch}           
           \and
           M. H\"{u}tter \at Materials Technology, Mechanical Engineering, Eindhoven
University of Technology, 5600 MB Eindhoven, The Netherlands
           }

\date{Received: 11. 12. 2010 / Published online: 11. 2. 2012}

\sloppy

\maketitle

\begin{abstract}
The macroscopic mechanical properties of colloidal particle gels
strongly depend on the local arrangement of the powder
particles. Experiments have shown that more heterogeneous
microstructures exhibit up to one order of magnitude higher
elastic properties than their more homogeneous counterparts at
equal volume fraction. In this paper, packings of spherical
particles are used as model structures to computationally
investigate the elastic properties of coagulated particle gels
as a function of their degree of heterogeneity. The discrete
element model comprises a linear elastic contact law, particle
bonding and damping. The simulation parameters were calibrated
using a homogeneous and a heterogeneous microstructure
originating from earlier Brownian dynamics simulations. A
systematic study of the elastic properties as a function of the
degree of heterogeneity was performed using two sets of
microstructures obtained from Brownian dynamics simulation and
from the void expansion method. Both sets cover a broad and to a
large extent overlapping range of degrees of heterogeneity. The
simulations have shown that the elastic properties as a function
of the degree of heterogeneity are independent of the structure
generation algorithm and that the relation between the shear
modulus and the degree of heterogeneity can be well described by
a power law. This suggests the presence of a critical degree of
heterogeneity and, therefore, a phase transition between a phase
with finite and one with zero elastic properties.

\end{abstract}

\section{Introduction}\label{intro}

Random sphere packings are ubiquitous model systems for the
study of the structural and mechanical properties of granular
matter or colloids. Geomechanics~\cite{Fakhimi_2007}, granular
flow~\cite{Rognon_2006} and mixing and segregation of granular
materials~\cite{Ottino_2000} are but a few examples. The
mechanical properties of granular systems depend on various
parameters such as the volume fraction~\cite{Zaccone_2007}, the
particle size distribution~\cite{Gardiner_2006}, material
properties as, for example, the particles' friction
coefficients~\cite{Silbert_2002} or adhesive
forces~\cite{Martin_2008}. Furthermore, as predicted
in~\cite{Cates_1999} and discussed in more detail in the
following, the mechanical properties strongly depend on the
microstructure, i.e., the local arrangement of the particles.

The influence of the microstructure on the mechanical properties
is often observed implicitly in experimental and computational
mechanical tests on structures differing in preparation history.
Macroscopic stress profiles, for example, where found to
strongly depend on the sample preparation procedure and thus its
microstructure~\cite{Atman_2005}. However, systematic
investigations of the mechanical properties as a function of the
microstructure are scarce, for two reasons: first, a systematic
study of the microstructure-dependent mechanical properties
requires the possibility of an unambiguous characterization of
the microstructural arrangement of the particles. Second, and
equally important, it relies on the possibility of a
reproducible generation of microstructures with distinct local
arrangements of the particles at constant volume fraction.

In~\cite{Schenker_pre_2009}, the concept of the degree of
heterogeneity (DOH) has been introduced, constituting a scalar
measure of the heterogeneity of the microstructural arrangement
of a sphere packing. Three distinct structure characterization
methods in conjunction with parameters in fit functions or
integrals were shown to allow for a clear quantification and
thus classification of the DOH of particle structures. In
contrast to distributional structure characterization methods,
such as the pair correlation function~\cite{Huetter_2000}, the
bond angle distribution~\cite{Huetter_2000}, the common
neighbors distribution~\cite{Schenker_2008} or the Minkowski
functionals~\cite{Huetter_2003}, the concept of the DOH does not
rely on the interpretation of distribution curves. This
quantifiability is seen as an advantage in view of a correlation
of the mechanical to the structural properties.

Experimentally, the reproducible control of the DOH of colloidal
microstructures with volume fractions between 0.2 and 0.6 is
achieved using direct coagulation
casting~\cite{Gauckler_1999,Tervoort_2004}, which is an \it in
situ \rm enzyme-catalyzed destabilization method. It allows for
the coagulation of electrostatically stabilized colloidal
suspensions to stiff particle structures by either shifting the
pH of the suspension to the particles' isoelectric point or by
increasing the ionic strength of the suspension. Shifting the pH
leads to ``more homogeneous'' microstructures through
diffusion-limited aggregation ($\mathrm{\Delta}$pH-method) while
increasing the ionic strength results in ``more heterogeneous''
microstructures via reaction-rate-limited aggregation
($\mathrm{\Delta}$I-method). These differences in heterogeneity
have been observed using various experimental techniques such as
diffusing wave spectroscopy~\cite{Wyss_2001}, static light
transmission~\cite{Wyss_2001} or cryogenic scanning electron
microscopy~\cite{Wyss_2002}.

\begin{figure}
\includegraphics[width=\linewidth]{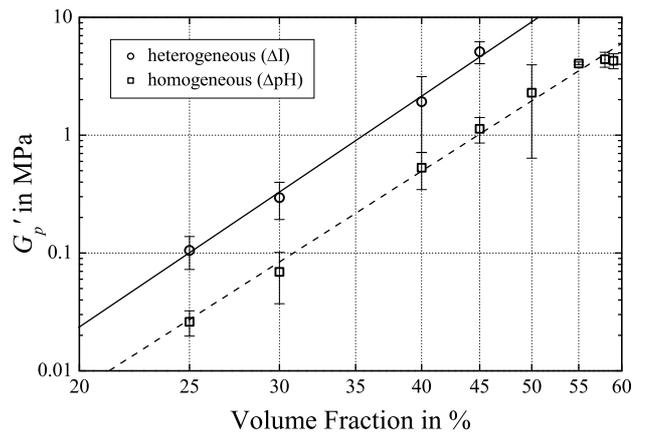} \caption{Experimental
plateau storage modulus $G'_p$ of alumina particle suspensions
(average particle diameter $\mathit{d_0}$~=~0.4~$\mu$m) formed
by the $\mathit{\Delta}$pH- and the $\mathit{\Delta}$I-method of
the direct coagulation casting process in dependence of the
volume fraction~\cite{Wyss_1_2005}.} \label{fig:fig1}
\end{figure}

Rheological and uniaxial compression experiments on coagulated
colloidal particle structures obtained by direct coagulation
casting have revealed that those with a more heterogeneous
microstructure have significantly higher elastic moduli than
their more homogeneous
counterparts~\cite{Balzer_1999,Wyss_1_2005,Wyss_2004}. The
rheological properties, which are the subject of the
computational part of this study, were investigated
experimentally using a Bohlin rheometer (Model CS-50, Bohlin
Instruments, Sweden) equipped with a measuring tool of
plate/plate geometry (rough surface, 25~mm plate diameter).
Oscillatory measurements were performed at a fixed frequency of
1~Hz with increasing strain amplitude. Figure~\ref{fig:fig1}
summarizes the measured plateau storage moduli $G'_p$, i.e. the
elastic shear modulus in the linear regime (small deformations),
of alumina particle suspensions (average particle diameter $d_0
= 0.4$~$\mu$m) destabilized by the $\mathrm{\Delta}$pH- and the
$\mathrm{\Delta}$I-method, respectively, as a function of the
volume fraction. For this system, approximately four times
higher elastic properties are measured for heterogeneous than
for homogenous microstructures at corresponding volume
fractions~\cite{Wyss_1_2005,Balzer_2_1999}.

A variation of the $\mathrm{\Delta}$pH-method allowing for a
controlled introduction of heterogeneities is the use of
alkali-swellable polymer particles. Small amounts of these
particles, 80~nm in diameter, are admixed to the suspended
powder particles under acidic conditions. The polymer particles
swell upon changing pH during the internal gelling reaction of
the direct coagulation casting process and enfold to 0.7~$\mu$m
in diameter, thereby rearranging the powder particles and thus
producing more heterogeneous microstructures. These more
heterogeneous samples exhibit much higher mechanical properties
in comparison to samples without polymer particles. In
particular, they present comparably high mechanical properties
as samples with heterogeneous microstructures produced by the
$\mathrm{\Delta}$I-method~\cite{Hesselbarth_2001,Hesselbarth_2000}.

In summary, strong evidence is given that the differences in
macroscopic mechanical properties of coagulated particle
suspensions are controlled by the differences in heterogeneity.
A yet unanswered question is how these microstructural
differences on the length scale of a few particle diameters can
have such a dramatic influence on the macroscopic mechanical
properties.

The aim of this study is to perform a systematic computational
analysis of the elastic shear properties of particle packings at
constant volume fraction of $\Phi = 0.4$ and to correlate these
properties with the structures' DOH.

\section{Materials and Methods}\label{MatAndMet}

In a first part of this section, the discrete element method,
which is the simulation method used throughout this study, is
introduced. Second, the methods, by which the two sets of
initial microstructures are obtained are briefly presented.
Then, the method and fit function that is used in order to
quantify the structures' DOH is explained and finally, the
simulation setup is presented.

\subsection{Discrete Element Method}

The discrete element method~\cite{Cundall_1979} is a numerical
method that allows simulating the motion of large numbers of
interacting particles. The simulations presented in this paper
are performed using the particle flow code in three dimensions
PFC$^{\mathrm{3D}}$~\cite{PFC_1995}, which is an implementation
of the discrete element method using spherical particles. One
calculation step consists in the determination of the total
force on each particle and the numerical integration of the
equations of motion using a central difference scheme. The force
calculation and the determination of the next positions and
velocities are then repeated until the end of the simulation.

Our model comprises a linear elastic contact law between
particles, particle bonding and damping. The linear elastic
contact law relates the contact forces acting on two particles
in contact linearly to the relative displacement between the
particles. In PFC$^{\mathrm{3D}}$ a soft-contact approach is
used, wherein the rigid particles are allowed to overlap at
contact points. The magnitude of the normal contact force $F_n$
is given by

\begin{equation}
 F_n = k_n U_n,
 \label{eq:sim_n_stiff}
\end{equation}

\noindent where $k_n$ denotes the normal stiffness and $U_n$ is
the overlap distance ($U_n > 0$). The shear stiffness $k_s$
relates incremental displacements in shear direction
$\mathrm{\Delta} U_s$ to the shear contact force
$\mathrm{\Delta} F_s$ via

\begin{equation}
 \mathrm{\Delta} F_s = k_s \mathrm{\Delta} U_s.
 \label{eq:sim_s_stiff}
\end{equation}

\noindent The linear elastic contact law is thus parameterized
by the normal and shear stiffnesses.

The so-called contact bonds used in PFC$^{\mathrm{3D}}$
essentially extend the linear elastic contact law,
equations~(\ref{eq:sim_n_stiff}) and~(\ref{eq:sim_s_stiff}), to
particles that are separated, i.e., to particles without overlap
($U_n < 0$). These bonds are characterized by a normal and shear
strengths, $F^B_n$ and $F^B_s$. A bond breaks if either the
normal or the shear bond strength is exceeded in normal and
shear direction, respectively.

Energy dissipation is simulated via a local, non-viscous damping
term added to the equations of motion~\cite{Brown_1987}. This
damping force $\vec{F_d}$ is characterized by the damping
coefficient $\alpha$ and is given componentwise (index~$i \in
\{1, 2, 3\}$) by

\begin{equation}
 F_{d,i} = -\alpha\; |F_i|\; \mathrm{sign}(v_i),
 \label{eq:damp}
\end{equation}

\noindent where $\vec{F}$ is the force acting on a particle,
$\vec{v}$ is the particle's velocity and

\begin{equation}
 \mathrm{sign}(x) = \left\{ \begin{array}{rl}
 +1,\;  & \mathrm{if}\;  x > 0\\
 -1,\;  & \mathrm{if}\;  x < 0\\
 0,\;  & \mathrm{if}\;  x = 0 \end{array} \right..
\end{equation}

\noindent The damping force is thus proportional to the force
acting on the particle and indeed leads to energy-dissipation
since $\vec{F_d}\cdot\vec{v} < 0$. In particular, only
accelerating motion is damped~\cite{PFC_1995}. An analogous
expression to Eq.~(\ref{eq:damp}) is used to describe the
damping of the particles' rotational motion.

\begin{table}
\caption{Simulation parameters.} \label{tab:sim_1}
\begin{tabular}{lll}
\hline\noalign{\smallskip}
Parameter & Symbol & Value  \\
\noalign{\smallskip}\hline\noalign{\smallskip}
Number of particles & $N$ & $8\,000$ \\
Particle radius & $r_0$ & 0.25~$\mu$m \\
Normal particle stiffness & $k_{n}$ & 50 -- 150 N/m \\
Shear particle stiffness & $k_{s}$ & 5 -- 15 N/m \\
Normal bond strength & $F^B_n$ & 10$^{-4}$ N \\
Shear bond strength & $F^B_s$ & 10$^{-6}$ N \\
Damping coefficient & $\alpha$ & 0.9 \\
Friction coefficient & $\mu$ & 0.0 \\
Volume fraction & $\Phi$ & 0.4 \\
Particle density & $\rho$ & $3\,690$ kg/m$^{3}$ \\
\noalign{\smallskip}\hline
\end{tabular}
\end{table}

In conclusion, the simulation model is characterized by five
microscopic parameters: the stiffness of the particles and the
bond strength both for normal and shear direction, and the
damping coefficient. These parameters, in the following referred
to as microparameters, are compiled in Table~\ref{tab:sim_1}. In
particular, the normal to shear particle stiffness ratio
$k_{n}/k_{s}=10$ is kept constant and a value of 0.9 is chosen
for the damping coefficient, which corresponds to the values
used in previous
studies~\cite{Schenker_pre_2009,Schenker_vem2_2009}. In
PFC$^{\mathrm{3D}}$ the frictional term between two particles is
superseded in the presence of a bond. However, it may become
active in the case of larger deformation when unbonded particles
come into contact. In order to prevent such contacts to result
in friction, a friction coefficient $\mu=0$ was used. It has to
be mentioned that the present paper does not include a detailed
sensitivity study of every simulation parameter. Here, the focus
is on the influence of the microstructural differences on the
mechanical properties. This influence can \it a priori \rm also
be observed with other choices of parameters as long as a unique
set is used for all structures.

\subsection{Initial Microstructures}

Two sets of microstructures are used as initial particle
configurations for the simulation of the elastic properties. A
first set originates from previous Brownian dynamics (BD)
simulations~\cite{Huetter_2000,Huetter_1999} where the
coagulation of electrostatically stabilized colloidal
suspensions to stiff particle structures was simulated. The
presence and depth of a secondary minimum in the inter-particle
potential, described by the Derjaguin-Landau-Verwey-Overbeek
theory~\cite{Russel_1989}, was shown to account for the
variations in the DOH of the resulting particle structures and
is adjusted via the surface potential $\Psi_0$. For $\Psi_0 =
0$~mV, the electrostatic double layer repulsion is zero and the
inter-particle potential is only given by the attractive van der
Waals potential. For the set of simulation parameters used
in~\cite{Huetter_2000,Huetter_1999}, a secondary minimum appears
for $\Psi_0 \geq 12$~mV and an energy barrier between the local
maximum and the secondary minimum emerges. For $\Psi_0 = 15$~mV,
a repulsive barrier of $5.65\,k_B T$ is present. The model
further contains the frictional Stokes' drag force and a random
Brownian force caused by the suspending liquid. In this paper,
BD-microstructures generated with $\Psi_0 = 0, 12, 13, 14$~and
15~mV are used, exhibiting an increasing DOH with increasing
$\Psi_0$~\cite{Schenker_pre_2009}.

In particular, the most and least heterogeneous
BD-microstructures have been shown to nicely correspond to
experimental silica structures using the pair correlation
function~\cite{Wyss_2002}. These two structures are thus used
here to calibrate the microparameters in order for the
simulations to reproduce the experimental values given in
Fig.~\ref{fig:fig1} (cf.~Sec.~\ref{sec:sec_31}).

The second set of initial microstructures is obtained using the
void expansion method (VEM) presented
in~\cite{Schenker_vem_2009}. In contrast to BD, which simulates
the physical processes during coagulation according to
established laws and methods, VEM is a purely stochastic method
inspired by the experimental generation of heterogeneous
microstructures using alkali-swellable polymer particles. This
method employs so-called void-particles that mimic the polymer
particles. The void-particles, having a small initial diameter,
are randomly placed into the simulation box containing the
structure-particles. The core part of VEM is the repeated
increase in the void-particles' diameter. During this procedure,
the structure-particles are rearranged and pushed into contact.
Finally, the void-particles are deleted. As shown
in~\cite{Schenker_vem2_2009}, VEM allows generating
microstructures presenting a broad range of DOH, which is
controlled by the void- to structure-particle number ratio. In
particular, VEM-microstructures cover an approximately 20\%
larger range of DOH, slightly shifted to higher values than the
BD-microstructures. In this study, the number of void-particles
$N_V$ ranges between $1\,000$ and $16\,000$. The void-particles'
normal and shear stiffness used in VEM are $k_{n,V}=10^{2}$~N/m
and $k_{s,V}=10^{-2}$~N/m, respectively.

All structures are submitted to the same relaxation procedure
before undergoing the simulated mechanical testing. Bonds
between neighboring particles having a center-to-center
separation smaller than $d_{\epsilon} = (1 + \epsilon) 2 r_0$
with $\epsilon = 0.01$ are installed. Then, several calculation
steps are performed, where, after each step, the translational
and rotational velocities are set to zero. This allows for an
efficient reduction of particle overlaps and thus of the
internal strain energy in the structures without significant
changes in the particle positions. The relaxation process is
aborted as soon as the mean unbalanced force in the structure
has decreased to a constant value.

The structures investigated in this study have equal volume
fraction of $\Phi=0.4$, consist of $N=8\,000$ monodispersed
spherical particles having a radius $r_0=0.25$~$\mu$m and are
contained in a cubic simulation box with periodic boundaries and
edge length $L_{box}$ given by

\begin{equation}
 L_{box}=r_0\left(\frac{4N\pi}{3\Phi}\right)^{1/3}.
 \label{eq:sim_edgelength}
\end{equation}

\noindent The structures used in this study are provided as
electronic supplementary material (Online Resources 1-12).

\subsection{Degree of Heterogeneity}

In~\cite{Schenker_pre_2009}, three distinct methods allowing for
a quantification of the DOH using scalar measures were
introduced. Here, the DOH is characterized by means of the
cumulative pore size distribution $P(r_P>r)$ estimated using the
exclusion probability~\cite{Torquato_1990}.
Equation~(\ref{eq:sim_errFxn}) was shown to nicely fit
$P(r_P>r)$ of both the VEM- and the
BD-microstructures~\cite{Schenker_pre_2009,Schenker_vem2_2009}.

\begin{equation}
P(r_P > r) = 1 - \mathrm{erf} \left( \frac{r/r_0 - b}{a
\sqrt{2}} \right).
 \label{eq:sim_errFxn}
\end{equation}

\begin{figure}
  \includegraphics[width=\linewidth]{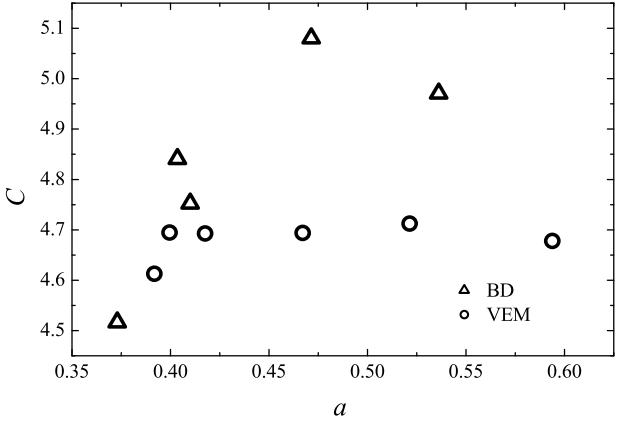}
\caption{Degree of heterogeneity as measured by parameter $a$ as
a function of the coordination number $C$ for all structures
used in this study.} \label{fig:fig2}
\end{figure}

The DOH is quantified by the width of the error function
measured by parameter $a$. The values of $a$ for the various
microstructures are summarized in Table~\ref{tab:sim_Gprime}.
Parameter $b$ represents the location of the maximum of the
underlying Gaussian distribution, i.e., the most probable pore
to particle radius ratio.

Additionally, the structures' coordination numbers $C$ are given
in~Table~\ref{tab:sim_Gprime} and plotted in Fig.~\ref{fig:fig2}
against the DOH-parameter $a$. This figure shows that the DOH
and the coordination number are to a large extent independent.

\subsection{Simulation Setup}

\begin{figure}
  \includegraphics[width=\linewidth]{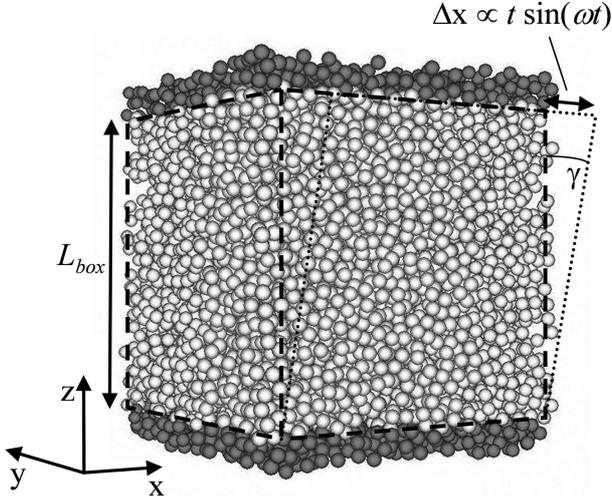}
\caption[Shear simulation setup]{Rheological simulation setup.}
\label{fig:fig3}
\end{figure}

The simulation setup is schematically shown in
Fig.~\ref{fig:fig3}. The structure (white particles) is placed
between periodic boundaries in x- and y-direction. In
z-direction, the boundary conditions are imposed via three
additional particle layers (gray), obtained from the periodic
repetition of the structure. The lower layers, representing the
lower plate in the shear experiment, are immobile whereas the
upper layers shear the sample periodically along the x-axis with
a fixed frequency of 1~Hz and constant volume. The shear
amplitude is increased after each oscillation. In particular,
ten oscillations with a linearly increasing deflection of the
upper plate up to a maximum deflection of 0.1\%~$L_{box}$ are
performed. Spherical regions, so-called measurement
spheres~\cite{PFC_1995}, are defined in different heights
monitoring the mean stress tensor in their respective region.
The absolute value of the complex modulus $G$ is obtained by
dividing the peak shear stress value with the actual peak strain
of each oscillation. The storage modulus $G'$ is then calculated
by multiplication with the cosine of the phase angle between
excitation and response, which corresponds to the evaluation of
$G'$ according to~\cite{Yanez_1996}. In total, the simulations
are performed six times for each structure. The shear plate's
normal vector is successively oriented along the x-, y- and
z-axis and, for each orientation, the structure is sheared along
the two remaining axes.

\section{Results and Discussion}

\subsection{Calibration of the Model} \label{sec:sec_31}

\begin{figure}
  \includegraphics[width=\linewidth]{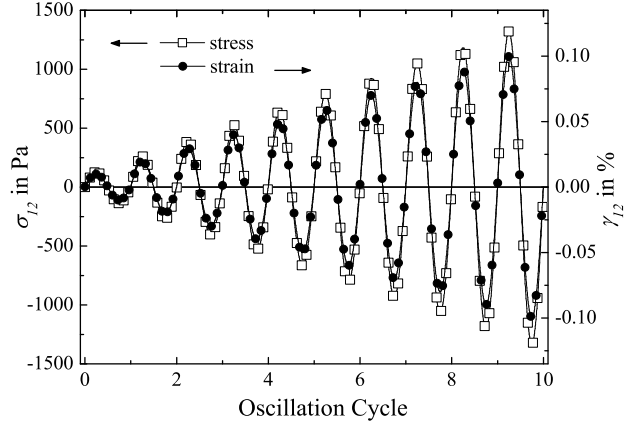}
\caption{Stress $\mathit{\sigma_{12}}$ (open squares, left
scale) and strain $\mathit{\gamma_{12}}$ (full circles, right
scale) as a function of shear oscillation cycle.}
\label{fig:fig4}
\end{figure}

Typical stress and strain curves are shown in
Fig.~\ref{fig:fig4}, presenting the shear stress $\sigma_{12}$
(open squares, left scale) and the applied shear deflection
$\gamma_{12}$ (full circles, right scale) as a function of the
shear oscillation cycle. $\sigma_{12}$ increases linearly with
$\gamma_{12}$ and no phase shift between excitation and response
is observed (phase angle~$<10^{-4}\pi$). This confirms that the
simulations present a purely elastic behavior and the constant
ratio $\sigma_{12}/\gamma_{12}$ is identified with the plateau
storage modulus $G'_p$.

\begin{figure}
  \includegraphics[width=\linewidth]{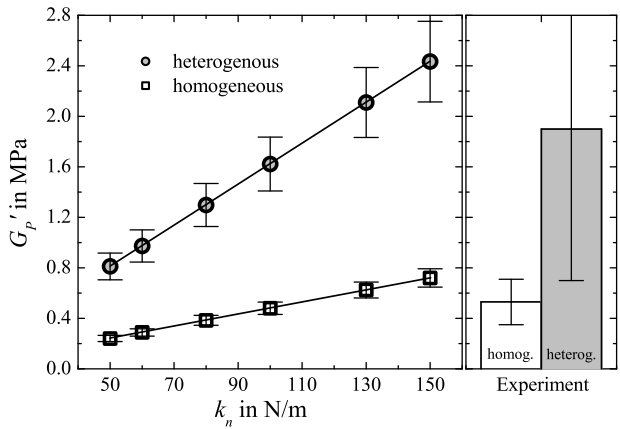}
\caption{Simulated plateau storage moduli $G'_p$ of the most and
least heterogeneous BD-microstructure as a function of the
particle normal stiffness $k_n$. Right: Experimental
$G'_p$-values for alumina particle structures (cf.\
Fig.~\ref{fig:fig1}). $\mathit{\Phi}$~=~0.4 for all
microstructures.} \label{fig:fig5}
\end{figure}

The particle stiffness $k_n$ has been calibrated using the most
homogeneous and the most heterogeneous BD-microstructures
($\Psi_0$~=~$0$~mV and $\Psi_0$~=~$15$~mV, respectively) using a
constant ratio $k_n/k_s = 10$. Its influence on the mechanical
properties of these two structures is shown in
Fig.~\ref{fig:fig5} (left graph). The experimental values are
shown on the right.

The simulated values for $G'_p$ quantitatively agree with the
experimental values within the experimental error for normal
stiffness values $k_n$ ranging between 50 and 150~N/m. For both
microstructures, a linear dependency is found between $G'_p$ and
$k_n$, which is expressed in a constant ratio between the
$G'_p$-values for the heterogeneous and homogeneous
microstructure as a function of $k_n$: $G'^{HE}_p/G'^{HO}_p
\approx 3.4$. This is in good agreement with the average
experimental $G'_p$-ratio of 3.6 at $\Phi$ = 0.4. For all
further simulations the particle normal stiffness is fixed to
$k_n=100$~N/m.

The inter-particle bond strength was strictly speaking not part
of the calibration process. In order to focus on the initial
elastic properties, its value was chosen high enough to prevent
any bond breakage for the deformations considered in this study.
The chosen normal bond strength $F^B_n=10^{-4}$~N, in
conjunction with the particle normal stiffness $k_n=100$~N/m
allows for a maximum particle separation distance of
$10^{-6}$~m, which corresponds to two particle diameters. Given
the maximum shear displacement of $0.001\,L_{box}$, where
$L_{box}$ is of the order of roughly 20~particle diameters, this
value is never exceeded.

\subsection{Elastic Properties as a Function of the DOH}

\begin{figure}
  \includegraphics[width=\linewidth]{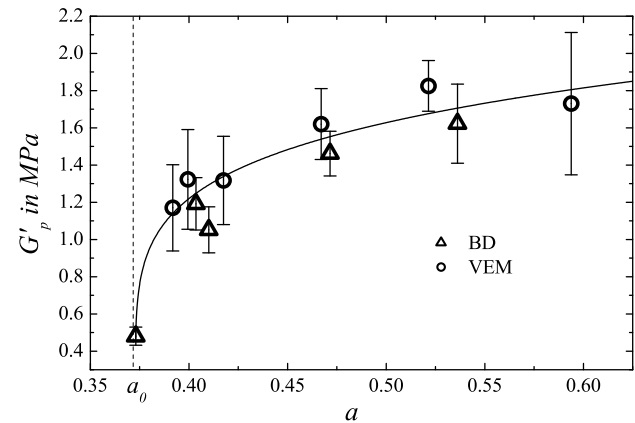}
\caption{Simulated plateau storage moduli $G'_p$ of the various
microstructures ($\mathit{\Phi}$~=~0.4,
$\mathit{r_0}$~=~0.25~$\mu$m) as a function of the DOH in terms
of parameter $a$ and power law fit (solid line).}
\label{fig:fig6}
\end{figure}

Using a particle normal stiffness of 100~N/m, the plateau
storage moduli of the various VEM- and BD-microstructure were
simulated. The resulting $G'_p$-values are summarized in
Table~\ref{tab:sim_Gprime} and are shown in Fig.~\ref{fig:fig6}
as a function of the structures' DOH expressed by parameter $a$.
Triangles and circles correspond to BD- and VEM-microstructures,
respectively. Each data point in Fig.~\ref{fig:fig6} corresponds
to one structure sheared in all six directions. The error bars
are thus reflecting the anisotropy of the structures.

Two conclusions can be drawn from Fig.~\ref{fig:fig6}: first,
the elastic properties present a clear dependence on the
microstructure's heterogeneity such that $G'_p$ increases for
increasing DOH-parameter $a$. Second, the behavior is
independent of the structure generation algorithm. Indeed, the
elastic moduli of the VEM- and the BD-microstructures agree
within the error bars for microstructures with comparable DOH.

\begin{table}
 \caption{DOH-parameter $a$~\cite{Schenker_pre_2009,Schenker_vem2_2009}, coordination
number $C$ and simulated plateau storage moduli $G'_p$ of the
various VEM- and BD-microstructures.} \label{tab:sim_Gprime}
 \begin{tabular}{lllll}
\hline\noalign{\smallskip}
 & Structure & $a$ & $C$ & $G'_p$ in MPa\\
\noalign{\smallskip}\hline\noalign{\smallskip}
\multirow{6}{*}{VEM} & $N_V=1\,000$ & 0.594 & 4.68 & $1.73 \pm 0.383 $\\
  & $N_V=2\,000$ & 0.521 & 4.71 & $ 1.83 \pm 0.136 $\\
  & $N_V=4\,000$ & 0.467 & 4.69 & $ 1.62 \pm 0.190 $\\
  & $N_V=8\,000$ & 0.418 & 4.69 & $ 1.32 \pm 0.237 $\\
  & $N_V=13\,000$ & 0.399 & 4.69 & $ 1.32 \pm 0.267 $\\
  & $N_V=16\,000$ & 0.392 & 4.61 & $ 1.17 \pm 0.231 $\\
\noalign{\smallskip}\hline\noalign{\smallskip}
 \multirow{5}{*}{BD} & $\Psi_0$ = 0 mV & 0.373 & 4.52 & $ 0.480 \pm 0.049 $\\
  & $\Psi_0$ = 12 mV & 0.404 & 4.84 & $ 1.19 \pm 0.141 $\\
  & $\Psi_0$ = 13 mV & 0.410 & 4.75 & $ 1.05 \pm 0.124 $\\
  & $\Psi_0$ = 14 mV & 0.472 & 5.08 & $ 1.46 \pm 0.120 $\\
  & $\Psi_0$ = 15 mV & 0.536 & 4.97 & $ 1.62 \pm 0.213 $\\
\noalign{\smallskip}\hline
 \end{tabular}
\end{table}

The solid line in Fig.~\ref{fig:fig6} represents a power law fit
given by

\begin{equation}
 G'_p \propto (a - a_0)^{\beta}, a \geq a_0,
 \label{eq:sim_plff}
\end{equation}

\noindent with $a_0 = 0.373 \pm 0.001$ (shown as dashed line in
Fig.~\ref{fig:fig6}) and $\beta = 0.187 \pm 0.032$. The use of
this fit function is inspired by percolation theory where a
power law scaling is found for the elastic properties as a
function of the volume
fraction~\cite{Benguigui_1984,Bergman_2002}. In view of this
analogy and based on the simulation data currently available,
this fit function suggests a phase transition with $a_0$ the
critical DOH. Below this value, the elastic properties are zero;
above this value, they increase for increasing $a$. A DOH-value
$a = 0.265$ below $a_0$ was found for the stabilized colloidal
microstructures with repulsive inter-particle
potentials~\cite{Schenker_pre_2009}. In the model used in this
work, this microstructure would indeed present negligible
elastic properties since there would be no bonds between the
particles.

It has to be mentioned that percolating structures presenting a
DOH below the critical value exist. Such structures, as for
example intermediate steps during the simulated coagulation
toward the most homogeneous microstructure using BD, are not
expected to have zero elastic properties. However, they have an
average coordination number significantly below the range
considered in this study and are therefore not included. Also,
it has to be emphasized that the obtained value for $a_0$ does
not solely rely on the data point possessing the lowest
$a$-value. Indeed, a power law fit without this data point
yields $a_0' = 0.360$, which is roughly 4\% below $a_0$.

The uncertainty of the value for the critical DOH $a_0$ roots in
fitting the power law (Eq.~\ref{eq:sim_plff}) to the simulated
plateau storage moduli of the samples studied here using error
propagation. The limits for $a_0$ do therefore not represent an
uncertainty obtained by analyzing a statistically significant
control sample. A more extensive study, involving a larger
number of microstructures possessing a DOH in the immediate
vicinity of $a_0$, i.e. between 0.35 and 0.4, would be required
in order to determine $a_0$ more accurately.

\section{Summary and Conclusions}
\label{Conclusions}

In this paper, the relation between the microstructure and the
elastic properties of packings of spherical particles was
investigated computationally using the discrete element method.
Two distinct sets of initial microstructures were subjected to
strain-controlled oscillatory shear simulations. The first set
originated from Brownian dynamics
simulations~\cite{Huetter_2000,Huetter_1999} where the physical
processes during coagulation were simulated. The second set was
obtained using the void expansion
method~\cite{Schenker_vem_2009}. In contrast to Brownian
dynamics simulations, the latter is a purely stochastic method.
Both sets cover a broad range of degrees of heterogeneity,
which, in this work, was characterized using the width of the
pore size distribution.

The simulations of the mechanical properties have been performed
using a discrete element model with five microparameters: the
particle stiffness and the bond strength in normal and shear
direction, respectively and damping. The inter-particle normal
and shear bond strengths were set to a high value in order to
prevent any bond breaking for the deformations applied during
the shear simulations. This allowed focussing on the initial,
purely elastic behavior found in experiment~\cite{Wyss_2004}.
The particle stiffness was calibrated using the two Brownian
dynamics microstructures presenting the highest and lowest DOH,
which correspond to experimental silica
structures~\cite{Wyss_2002}.

A quantitative agreement with experiment was achieved. For
particle normal stiffnesses ranging between 50~and 150~N/m, the
absolute values of the experimental elastic moduli are
reproduced, as shown in Fig.~\ref{fig:fig5}. In particular, the
ratio between the elastic moduli of the heterogeneous and the
homogeneous microstructures is in good agreement with the
experimental value. This latter result is particularly
remarkable, since the $G'_p$-ratio has not been subject of any
calibration but reflects the influence of the microstructures
and emphasizes the strong influence of the DOH independently of
the interaction potential. Indeed, the model used in this study
constitutes an important simplification with respect to
theoretical models such as the Derjaguin-Landau-Verwey-Overbeek
theory~\cite{Russel_1989} or the Johnson-Kendall-Roberts
theory~\cite{Johnson_1971}. The former gives the inter-particle
potential as the sum of the van der Waals attraction and the
electrostatic repulsion and the latter describes the adhesive
force between particles.

Using the calibrated model, the elastic properties of all
microstructures generated using the void expansion method and
Brownian dynamics were simulated. This investigation has
revealed a correlation between the elastic properties and the
structures' DOH. It has also shown that the elastic properties
are independent of the structure generation algorithm. Indeed,
for structures with comparable degrees of heterogeneity,
comparable plateau storage moduli were found. A power law fit
was shown to well reproduce the relation between the plateau
storage modulus and the DOH above a critical value, which
indicates a phase transition between a phase with finite and a
phase with zero elastic properties. This result is in nice
analogy to earlier findings~\cite{Benguigui_1984,Bergman_2002},
where the elastic properties were shown to exhibit a power law
dependence as a function of the volume fraction. This study thus
suggests an extension to percolation theory, where usually the
volume fraction constitutes the continuous variable presenting a
critical value. Here, the DOH, possessing the same critical
behavior as the volume fraction, has been identified as an
additional variable.

The results presented in this paper show that the degree of
heterogeneity is particularly useful in order to quantify and
characterize the heterogeneity of particle packings since it
allows for an unprecedented theoretical description of their
elastic properties. However, it does not directly lead to a
better understanding of the mechanisms, by which the elastic
properties increase for increasing heterogeneity. Such an
understanding must, in some way, include an intermediate step,
such as an analysis of the distribution of particle contacts
using the fabric tensor, for example~\cite{Madadi_2004}, or of
load-bearing substructures, as discussed
in~\cite{Schenker_2008}.

\begin{acknowledgements}
The authors would like to thank Hans J. Herrmann and Tomaso Aste
for their valuable help throughout this project.
\end{acknowledgements}

\bibliographystyle{spphys}
\bibliography{my_bib}

\end{document}